\documentstyle[12pt,epsfig]{article}

\newcommand{\mrm}[1]{\mbox{\rm #1}}

\newcommand{\beq}{\begin{equation}}
\newcommand{\eeq}{\end{equation}}
\newcommand{\nn}{\nonumber}
\newcommand{\bea}{\begin{eqnarray}}
\newcommand{\eea}{\end{eqnarray}}

\newcommand{\gsim}{\ \rlap{\raise 2pt\hbox{$>$}}{\lower 2pt \hbox{$\sim$}}\ }
\newcommand{\lsim}{\ \rlap{\raise 2pt\hbox{$<$}}{\lower 2pt \hbox{$\sim$}}\ }

\newcommand{\np}[1]{Nucl. Phys. {\bf #1}}
\newcommand{\pl}[1]{Phys. Lett. {\bf #1}}
\newcommand{\pr}[1]{Phys. Rev. {\bf #1}}
\newcommand{\prl}[1]{Phys. Rev. Lett. {\bf #1}}
\newcommand{\zp}[1]{Z. Phys. {\bf #1}}

\newcommand{\ptp}[1]{Prog. Theor. Phys. {\bf #1}}

\makeatletter
\if@twoside
\else \oddsidemargin 00pt \evensidemargin 00pt
\fi
\let\@eqnsel = \hfil

\expandafter\ifx\csname mathrm\endcsname\relax\def\mathrm#1{{\rm #1}}\fi
\@ifundefined{mathrm}{\def\mathrm#1{{\rm #1}}}{\relax}

\makeatother

\begin{document}

\thispagestyle{empty}
\null
\hfill FTUV/96-71, IFIC/96-80

\hfill hep-ph/9704211

\vskip 1.5cm

\begin{center}
{\Large \bf      
MASS OF TAU NEUTRINO IN $SO(10)$ GUTS
\par} \vskip 2.em
{\large		
{\sc G. Barenboim and M. Raidal
}  \\[1ex] 
{\it  Departament de F\'\i sica Te\`orica, Universitat 
de Val\`encia} and \\ 
{\it  IFIC, Centre Mixte Universitat 
de Val\`encia - CSIC} \\
{\it E-46100 Burjassot, Valencia, Spain} \\[1ex]
\vskip 0.5em
\par} 
\end{center} \par
\vfil
{\bf Abstract} \par
We investigate the allowed ranges of masses for an unstable tau neutrino 
in the context of $SO(10)$ GUT-s. In light of the new nucleosynthesis  
results we obtain that there is a narrow window for $m_{\nu_\tau}$ where 
the LEP, neutrino oscillation and nucleosynthesis data are compatible.
This window, which depends on the effective number of neutrinos 
contributing to nucleosynthesis, has important cosmological
consequences and will be tested by ongoing neutrino oscillation and LEP II 
experiments.
\par
\vskip 0.5cm
\noindent December 1996 \par
\null
\setcounter{page}{0}
\clearpage

The question of understanding the  smallness of the ordinary
 neutrino masses, $m_{\nu_e}$, $m_{\nu_\mu}$ and 
$m_{\nu_\tau},$  relies on different solutions in the context of
extended gauge structures of the Standard Model (SM).
The most attractive one, known as the
see-saw mechanism \cite{one}, has been conceived in $SO(10)$
grand unified theories (GUT) \cite{so}.
In these theories, the light neutrinos are massive Majorana particles and,
in general, mixed between flavour states.
Neutrino mixing implies that the generational lepton numbers,
like the electron number ($L_e$), muon number ($L_{\mu}$) 
and taon number ($L_{\tau}$), are not
conserved, giving rise to flavour changing processes like radiative decays 
of $\mu$ and $\tau$ leptons. 
Apart from these, another type of flavour changing decays 
of charged  leptons, 
containing only charged leptons and antileptons in the final states,
has been extensively searched for in experiments.
Taking into account the present experimental limits on 
muon decays \cite{pdb},
there is really little room remaining for the violation
of $L_\mu + L_e$. However, the present upper limits on the 
branching ratio  for rare tau decays 
provide  a possibility of violating   $L_e + L_\tau$ 
or $L_\mu + L_\tau$ at a much higher level.

The $Z^0$ width measurements at LEP 
have limited the  number of ordinary neutrino
species to three with an impressive accuracy. A consistent bound,
$N_\nu\lsim 3.04 $ \cite{kernan}, has been derived from studies of 
big bang nucleosynthesis (BBN) in the early Universe.
However,  recent analyses \cite{olive,fourd}
show an inconsistency between the standard BBN prediction and
observed abundances of primordial $^4$He and D, and  
give $N_{\nu}=2.1\pm 0.3$ \cite{fourd}
ruling out $N_{\nu}=3$ at 98.6\% C.L.
To solve the problem one has to conclude that one of the neutrinos, 
necessarily the tau neutrino, is unstable and, 
in order not to contribute as one neutrino specie to the BBN,
 its lifetime $\tau_{\nu_\tau}$ should be limited to 
\bea
\tau_{\nu_\tau}\lsim 1\;\; \mrm{s}.
\label{newone}
\eea
An additional strong requirement arises immediately for the neutrino mass. 
To avoid its background production via 
$e^+e^-\rightarrow \bar{\nu}_\tau \nu_\tau$ at BBN 
time its mass should satisfy $m_{\nu_\tau}\gsim{\cal O}(1)$ MeV.

In this letter we extend the analyses of Ref. \cite{four} and 
derive lower bounds on
the tau neutrino mass in general $SO(10)$ GUT models
taking into account laboratory limits on flavour changing processes 
and nucleosynthesis results. 
First we consider the case with $N_\nu=2$ effective neutrinos
contributing to BBN. Later on, we relax the requirement and allow $N_\nu$ 
to be around 3.
The obtained bounds depend on the mixing angle of tau neutrino as well
as on the bounds on the branching ratio of the process 
$\tau \rightarrow 3 \mu$ and, most strongly,
 on the mass of new neutral Higgs boson.
Therefore, they can be probed through  collider experiments and 
searches for neutrino mixings and other lepton number violating
processes. 

Among the two maximal continuous subgroups of  $SO(10),$ 
$SU(5)$ and $SU(2)_R \times SU(2)_L \times SU(4),$ 
which can appear in the symmetry breaking chain, 
only the latter one is viable phenomenologically. 
The Pati-Salam gauge group \cite{pasa} 
displays the left-right symmetry directly, implying that all quarks 
and leptons are assigned to left- and right-handed doublets.  
The Pati-Salam symmetry can break further into the 
left-right gauge group $SU(2)_R \times SU(2)_L \times U(1)_{B-L}$
which breaks down to the SM. 
This can be done in several steps, passing through different
symmetry groups or in only one step, directly \cite{lr}.  

Concerning the scalar sector, 
any model should contain at least
one representation $\Phi(10)$ necessary to generate Dirac masses 
for fermions and one representation $\Phi(126)$ necessary to
generate Majorana masses for the right-handed neutrinos and give rise to 
the see-saw mechanism. To avoid arbitrary complications we 
choose to work with a minimal model containing only one $\Phi(10)$ and
one $\Phi(126).$
The $\Phi(10)$  is a complex vector representation of Higgs field
whose content under the chiral decomposition  is given by
\bea
10 \; = \;\left( 2, 2, 1 \right) \; + \; \left( 1, 1, 6 \right). \nn
\eea
Therefore,
derived from a single $\Phi(10),$ there can be one or two 
Higgs doublets in our low energy theory.   
Since no unification of $m_b \approx m_\tau $ at the breaking
scale of the Pati-Salam subgroup can be achieved with only one
doublet \cite{des}, only the two doublet models are well motivated in
the context of GUT-s.   
What is relevant for our analyses is that triplet representations,
denoted by $\Delta_{R,L},$ exist
in both right- and left-handed sectors. 
They arise from the $\Phi(126)$ representation which decomposes as
\bea
126 \; = \; \left( 1, 1, 6 \right) 
 +  ( 3, 1, \overline{10})
 +  \left( 1, 3, 10 \right)  +  \left( 2, 2, 15 \right). \nn
\eea
The neutral component of 
the left-handed triplet mediates the non-diagonal neutrino decays which
we investigate in this work. 
As we work with the minimal scalar content, we are
going to have only one
 $\Delta_L$ and one $\Delta_R.$
However, our results
will be unchanged if more scalars would be added.

With this field content the effective Yukawa Lagrangian 
below the Pati-Salam breaking scale can be written as
\bea
\cal{L}_Y  & = &  \overline{\Psi}_{L} h \phi \Psi_R
 +   \overline{\Psi}_{L}\tilde{h} \tilde{\phi} \Psi_{R} +  \mbox{h.c.}
  \nn \\
& + & i \left(
 \Psi^T_{L} C \tau_2  \vec{\tau} \cdot f \vec{\Delta}_{L} \Psi_{L} 
 +  \Psi^T_{R} C \tau_2 \vec{\tau} \cdot f \vec{\Delta}_{R}
 \Psi_{R} \right)  +
 \mbox{h.c.},
\label{yuk}
\eea
where $\Psi_{L,R}$ denote the left- and right-handed lepton doublets,
respectively, $\phi$ a bidoublet of Higgs fields, 
$\tilde{\phi} = \sigma_2 \phi^*
\sigma_2,$ and 
$h$, $\tilde{h}$ and $f$ are matrices of Yukawa coupling
constants in generation space.
As required by phenomenology, the vacuum expectation value (vev)
$\langle \phi \rangle $ gives masses to the charged leptons (and analogously 
to quarks) and
Dirac masses to the neutrinos, whereas a large  
$\langle \Delta^0_{R} \rangle $ leads to the see-saw mechanism for
neutrinos. The vev of the left triplet, 
$\langle \Delta^0_{L} \rangle, $ is strongly constrained due to its 
contribution to the $\rho$ parameter and can be taken to be zero.

Since our purpose 
is to study the possible degree of $L_e + L_\tau$ 
or $L_\mu + L_\tau$  violation,
we have to discuss experimental constraints on the Yukawa couplings
in Eq. (\ref{yuk}). The branching ratios of the processes
$\mu\rightarrow 3e$ and $\mu\rightarrow \gamma e$ are constrained 5-6
orders  of magnitude more stringently than the ones of 
other flavour changing decays of leptons \cite{pdb}. 
 It follows from the former process that $f_{e\mu}$ is negligible 
compared with the other values of $f_{ij}$ and can be neglected \cite{moha2}.
The decay $\mu\rightarrow \gamma e$ sets constraints on the combination 
$f_{e\tau}f_{\tau\mu}$ of the couplings and imply that one of them can be
neglected if the other is assumed to be sizable. Similar argumentation 
applies also to the Yukawa couplings $h_{ij}.$ Mixing angles  of
the left-handed neutrinos are given with good accuracy (up to  
corrections of order $\langle\phi\rangle/\langle\Delta_R\rangle$)
by the matrix $h,$ and non-observation of $\mu\rightarrow \gamma e$ 
constrains the $\nu_e\nu_\mu$ and  $\nu_\tau\nu_e$ or $\nu_\tau\nu_\mu$
mixings very strongly. 
On  phenomenological
grounds, these constraints are equivalent to imposing the approximate global 
$U(1)_{\tau + \mu}$ or $U(1)_{\tau + e}$ symmetries\footnote{
To be mathematically rigorous, one has to impose an exact global $U(1)$
symmetry and double the number of triplet Higgs multiplets. 
It was shown in Ref. \cite{four} that for all phenomenological
purposes these two approaches are equivalent.}.   
We will work with the most natural model with the 
$U(1)_{\tau + \mu}$ global symmetry and comment later on the implications
of imposing the $U(1)_{\tau+e}.$
In this limit, the electron generation 
separates entirely  from the $\mu$ and $\tau$ generations 
and the decay $\mu \rightarrow 3e$ 
is forbidden. The Yukawa matrices in Eq.(\ref{yuk}) take the form
\bea
f  \equiv \pmatrix{ f_{ee}  & 0 & 0 
 \cr
0& f_{\mu\mu}  & f_{\mu\tau}  \cr
0 & f_{\tau\mu} 
 & f_{\tau\tau}  \cr}, \nn
\eea
and similarly for $h$ and $\tilde{h}$.

The physics we are interested in comes from the left-handed triplet 
sector of the theory. Since the $\Delta_{L}$
fields do not take part in the Higgs mechanism,  we can ignore 
the couplings of the type  $\Delta_{L} \phi
\Delta_{R}^+ \phi^+$ in the Higgs potential.  
In this case the $\Delta_{L}$ and $\Delta_{R}$ remain
unmixed states. In the basis 
where all leptons are mass eigenstates the relevant Yukawa Lagrangian 
is given by
\bea
\cal{L}_Y   =  \nu_L^T F^\prime C^{-1} \Delta^0_{L} \nu_L  + 
\nu_L^T F^{\prime \prime} C^{-1} \Delta^+_{L} L_L   +  
L_L^T F C^{-1} \Delta^{++}_{L} L_L  +  \mbox{h.c.}, \nn
\eea
where $\nu = \left( \nu_2 , \nu_3 \right)$, $L = \left(
\mu , \tau \right)$ and $F$, $F^\prime$ and $F^{\prime \prime}$ 
are $2\times2$ matrices related to each other as 
$ F K^T = F^{\prime \prime},$ and
$ K F K^T = F^\prime.$
Here $K$ is the leptonic CKM matrix in the left sector
involving only the neutrino mixing angle, $\theta_{\mu\tau}$, 
which can be  measured
in neutrino oscillation experiments. 
For large neutrino masses the present limit on the mixing angle is
$ 3\cdot 10^{-2}$ \cite{pdb}. Therefore, we denote always 
$\nu_{2,3}\equiv\nu_{\mu,\tau}.$ 
This bound will be updated, 
if the angle will not be measured, 
already this year by CHORUS and NOMAD experiments 
which have been taking data since 1994 and will 
achieve the designed sensitivity  
of $\theta_{\mu\tau} \leq 7\cdot 10^{-3}$ \cite{fourb}. 
The planned Fermilab E803 and NAUSICAA  experiments will have the potential  
to measure also $\theta_{e\tau}$ with the same sensitivity or better, and 
$\theta_{\mu\tau}$ at the level of $10^{-3}$  \cite{naus}.

Our results will explicitly depend on the masses of the left triplet Higgs 
bosons which are set by some new symmetry breaking scale.   
However, since $\Delta_{L}^0$ ,
$\Delta_{L}^+$ and $\Delta_{L}^{++}$ belong to the same 
$SU(2)_L$ multiplet, their mass difference should be of the
order of the $SU(2)_L$ breaking scale, i.e., a few hundred GeV at 
most. Indeed, in the class of $SO(10)$ GUT  models what 
we consider,  $M_{\Delta_{L}^0}$,
$M_{\Delta_{L}^+}$ and $M_{\Delta_{L}^{++}}$ are not independent. 
Starting with the most general form of the Higgs potential involving
the bidoublet and the 
triplets one can show that the following relations hold 
\cite{mopal} 
\bea
M_{\Delta_{L}^{++}}^2 = 
M^2_{\Delta_{L}^0} \left( 1 + 2 \alpha \right) , \;\;
M_{\Delta_{L}^{+}}^2  = M^2_{\Delta_{L}^0} \left( 1 +  \alpha
\right) , \nn
\eea
where $\alpha$ is a dimensionless combination
of the Higgs potential parameters. 
Experimentally the values of  $\alpha$ can be bounded from  
the measurements of the parameter 
$\rho = 1 + \rho_\theta + \rho_\Delta, $ 
where $\rho_\theta$ is a
correction  due to the mixing of $Z^0$ with 
a new neutral gauge boson
(which we are neglecting here) and $\rho_\Delta$ comes from
the $\Delta_L$-loop contribution to the $Z^0$ and $W^{\pm}$ mass. It is
given by \cite{eleven}
\bea
\rho_\Delta = \frac{G_F}{4 \sqrt{2} \pi^2} \left[ f_{(\Delta_{L}^0,
\Delta_{L}^+)} + 
f_{(\Delta_{L}^+,\Delta_{L}^{++})} 
\right] \equiv \frac{3 G_F}{8 \sqrt{2} \pi^2}
\Delta m ^2 , \nn
\eea
where
$ f_{(x,y)}=M_x^2 +M_y^2 - 2 M_x^2 M_y^2 \ln (
M_y^2/M_x^2 )/(M_y^2-M_x^2). $
Studies of the new contributions to the $\rho$ parameter 
have been settled the upper bounds  
$\Delta m^2  \leq (76 \; \mbox{GeV} )^2, $ 
$(98 \; \mbox{GeV} )^2,$ $ (122 \; \mbox{GeV} )^2$ \cite{pdb} 
for the SM Higgs masses $m_H = 60,$ 300 and 1000 GeV, respectively, at  
90$\%$ C.L.
The obtainable lower bounds on $\alpha$ depend strongly on 
$M_{\Delta_{L}^0}$ and $m_H,$ and are presented 
in Table 1. The present mass limit
$M_{\Delta_{L}^0} \gsim 45$ GeV, which derives from the LEP 
invisible $Z^0$ width measurement,
will improve at LEP II and Next Linear Collider (NLC) up to
the representative values 80 GeV and 250 GeV, respectively.

Let us now turn to the decays of $\nu_\tau,$ which can have the following 
 modes
\bea
 \nu_\tau \;  \longrightarrow  \;  \nu_\mu 
\gamma \, , \, 3 \nu_\mu \, . 
\label{decay}
\eea
The first, the radiative decay mode, is highly suppressed \cite{rad}
and is not useful in satisfying the constraint (\ref{newone}).
Therefore, we are left with the decay
$ \nu_\tau   \rightarrow   3 \nu $  mediated by the 
$\Delta^0_{L}$ exchange. The effective hamiltonian for this process
is given by
\bea
H = \frac{ G_{\nu_\tau}}{\sqrt{2}} \overline{\nu}_\mu \gamma^\lambda 
\left( 1 - \gamma^5 \right) \nu_\mu \, \overline{\nu}_\mu  \gamma_\lambda
 \left( 1 - \gamma^5 \right) \nu_\tau  +  \mbox{h.c.}, \nn
\eea
where 
$G_{\nu_\tau} = \sqrt{2}\left(
f_{\mu \mu} + 2 \theta_{\mu\tau} f_{\mu\tau} \right) \left( f_{\mu\tau} -
\theta_{\mu\tau} \left( f_{\mu\mu} - f_{\tau\tau} \right)\right)
/(4 M^2_{\Delta_{L}^0}). $
The calculation of  $\nu_\tau$ lifetime is
straightforward  and gives
$\tau_{\nu_\tau}^{-1} = 2 G_{\nu_\tau}^2 m_{\nu_\tau}^5/
(192  \pi^3).$
Using  the constraint (\ref{newone}) this yields
a lower bound 
\bea
m_{\nu_\tau}\gsim  
\frac{0.11 \;\mrm{MeV}\; 
\left(M_{\Delta_L^0}\mbox{GeV}^{-1}\right)^\frac{4}{5}}
{\left[\left(
f_{\mu \mu} + 2 \theta_{\mu\tau} f_{\mu\tau} \right) \left( f_{\mu\tau} -
\theta_{\mu\tau} \left( f_{\mu\mu} - f_{\tau\tau} \right)\right)
\right]^\frac{2}{5} } ,
\label{constr}
\eea
which 
depends explicitly on the bounds on $\Delta_L^0$ mass,
$\theta_{\mu\tau}$ and the Yukawa coupling constants $f_{ij}.$  
It is important to notice that even if $
f_{\mu \tau} = 0$ (i.e, even if there is no 
$\tau \rightarrow 3 \mu $ decay), 
  $\nu_\tau$ can decay due to the neutrino mixing.

In order to obtain numerical estimates on $m_{\nu_\tau}$ 
we have to consider constraints on the triplet Yukawa couplings.
Most generally, vacuum stability requires
$\mid f_{ij}\mid \leq 1.2 $ \cite{nine}, which is also the only 
available bound on $f_{\tau\tau}.$ From the extra contribution to 
$(g-2)_\mu$ the following bound has been established \cite{moha2}
\bea
| f_{\mu\mu}| \lsim 0.25\cdot 10^{-2}\sqrt{1+2\alpha}
\;\mrm{GeV}^{-1} M_{\Delta_L^{0}}.
\label{flim1}
\eea  
In the presence of the non-diagonal coupling $f_{\mu \tau},$ the 
decay $\tau \rightarrow 3 \mu$ mediated by  $\Delta_{L}^{++},$
is allowed. 
Analogously to the previous case,  its branching ratio is given by 
\bea
B(\tau \rightarrow 3 \mu)=\frac{f_{\mu \mu}^2 f_{\mu\tau}^2}
{4 \Gamma_\tau M_{\Delta_{L}^{++}}^4} \frac{m_\tau^5}{192 \pi^3}, \nn
\eea
where $\Gamma_\tau$ is the total width of the decaying lepton.
The present experimental  limit  $B(\tau \rightarrow 3 \mu) \leq
1.9 \cdot 10^{-6}$ \cite{argus} leads to the bound
\bea
|f_{\mu\mu}f_{\mu\tau}|\lsim 7.6\cdot 10^{-8} (1+2\alpha)\;\mrm{GeV}^{-2} 
M_{\Delta_L^{0}}^2.
\label{flim2}
\eea
All these constraints will improve considerably at NLC which will
measure the diagonal couplings, including $f_{\tau\tau},$ 
 with the same sensitivity as the off-diagonal ones \cite{nlc}. 
To obtain as conservative bounds on the neutrino mass 
as possible, we have used the 
maximum values of $\alpha$ in Table 1 and  assumed that the Yukawa
coupling constants in Eq. (\ref{constr}) add up constructively.
The most stringent constraint appears when the process
$\tau \rightarrow 3 \mu $ is forbidden and becomes 
somewhat relaxed if the decay $\tau \rightarrow 3 \mu $ is allowed.  
For the maximally allowed $f_{\mu\tau}$ we obtain
\bea
m_{\nu_\tau}\gsim 9 \;\mrm{MeV},
\label{bound1}
\eea 
which is the absolute lower bound on 
tau neutrino mass in the models we consider. 
Should CHORUS and NOMAD experiments show negative results,
the bound will rise to $m_{\nu_\tau}\gsim 14$ MeV.

These are  strong constraints, leaving a quite narrow window for the neutrino 
mass up to the current laboratory limit 
$m_{\nu_\tau}\lsim 18$ MeV \cite{aleph}.
On the other hand, a $\nu_\tau$ in this mass range 
has important cosmological consequences: 
it is  required by COBE anisotropy measurements 
to obtain the correct amount of 
cold dark matter in the Universe \cite{cobe}.  
It follows from Eq. (\ref{constr})  that
the bounds depend quite weakly on changes in the constraints on 
$\tau_{\nu_\tau}$ 
and $\theta_{\nu\tau}$ and relatively strongly on changes of $M_{\Delta_L^0}.$ 
If $\Delta_L^0$ will not be discovered in the running LEP II experiments    
the bound (\ref{bound1}) will rise up to 21 MeV,  
closing the allowed gap for tau neutrino
mass. This would lead into difficulties to understand the BBN
result, $N_\nu=2.1,$ in $SO(10)$ GUT-s. 

If we relax the constraint (\ref{newone}) and allow the possibility 
that three neutrinos contribute to BBN,
then the tau neutrino, provided its mass exceeds 100 eV, 
should be unstable and 
decay via the neutral Higgs exchange at some later time.
Most generally, from constraints of 
the mass density of the Universe \cite{uus}
one  obtains the bound \cite{eight,four} 
\bea
\tau_{\nu_\tau} \leq 8.2 \cdot 10^{31} \; \mbox{MeV}^{-1} \left(
\frac{100 \; \mbox{keV}}{m_{\nu_\tau}} \right)^2 .
\label{uno}
\eea
More stringent bounds can be derived from considerations of
galaxy formation \cite{six} but, as they are model dependent,
we are not going to treat them here. 
It follows from Eq. (\ref{uno}) that 
\bea
m_{\nu_\tau}\gsim 
\frac{3.1\cdot 10^{-2}\;\mrm{keV}\;
\left(M_{\Delta_L^0}\mbox{GeV}^{-1} \right)^
\frac{4}{3}}
{\left[\left(
f_{\mu \mu} + 2 \theta_{\mu\tau} f_{\mu\tau} \right) \left( f_{\mu\tau} -
\theta_{\mu\tau} \left( f_{\mu\mu} - f_{\tau\tau} \right)\right)
\right]^\frac{2}{3} } ,
\label{constr2}
\eea
implying $m_{\nu_\tau}\gsim 48$ keV
(which increases up to $m_{\nu_\tau}\gsim 102$ keV with 
 the expected CHORUS sensitivity)
for the most conservative situation.
In this case, recent BBN calculations using the full Boltzmann
equation give bounds in the range  
$m_{\nu_\tau}\lsim 0.1$ to 0.4 MeV \cite{kaw} 
for the Majorana neutrino mass.
This range will be tested by LEP II and forthcoming neutrino 
oscillation experiments or by the NLC  (see Table 1)
where a sensitivity of $m_{\nu_\tau}\gsim 0.5 $ MeV will be achieved.

It appears to be difficult to avoid the obtained bounds in $SO(10)$ GUT-s.
Enlargering the scalar sector may somewhat increase the effective coupling
$G_{\nu_\tau}$ but, due to the experimental constraints (\ref{flim1}) 
and (\ref{flim2}), no sizable effects can be achieved. 
Since there is no Yukawa couplings to singlets, adding new singlet neutrinos 
will not open any new decay channels. The possibility of adding 
new generations 
with light standard neutrinos is excluded by LEP.
Finally, one cannot suppress the triplet Yukawa couplings and 
force the right-handed neutrinos to be very light (no see-saw mechanism)
since the left-right mixing is known to be negligible and 
the unique decay mode of $\nu_\tau$ to three light neutrinos,
left- or right-handed, will be strongly suppressed.
A possibility to relax somewhat these bounds is to 
assume that, unlike the situation in quark sector, 
$\nu_e$ and $\nu_\tau$ are strongly mixed, implying a global 
$U(1)_{\tau+e}$ symmetry.  
However, this unnatural situation has a very little breathing space
and will become as restrictive as the $U(1)_{\tau+\mu}$ case after
the new oscillation data will be available.  Certainly,
in the case of larger gauge groups, e.g. $E_6,$ new decay channels occur.

To summarize, in $SO(10)$ GUT-s
the BBN result $N_\nu=2.1$ yields  
a lower bound 9 MeV (14 MeV if CHORUS and NOMAD
experiments will show negative results) on tau neutrino mass. 
Most importantly, this possibility will be tested in ongoing LEP II 
experiments. If more than two effective 
neutrino species  contribute to BBN, the presently allowed     
region for unstable tau neutrino mass, 0.05-0.4 MeV,  
will be probed by planned neutrino oscillation and collider experiments.
The presently allowed  mass ranges
are of great interest because of their important cosmological consequences.

We are indebted to M.C. Gonzalez-Garc\'{\i}a, C. Jarlskog, K. Kainulainen, 
E. Roulet, A. Santamar\'{\i}a and J.W.F. Valle for stimulating discussions.
G.B. acknowledges the Spanish Ministry of
Foreign Affairs for a MUTIS fellowship, whereas M.R.
thanks  the Spanish Ministry of Education for a post-doctoral fellowship.
This work was supported by CICYT, under grant AEN 96/1718.

\newpage

\subsection*{Table caption}
\begin{itemize}
\item[Table 1.] 
Upper bounds on values of $\alpha$ for different triplet and SM 
Higgs boson masses
\end{itemize}

\vspace*{3cm}

\begin{table}
\begin{center}
\begin{tabular}{||c|c|c|c||}\hline\hline
$M_{\Delta_L^0}$ (GeV) & $\alpha$ ($m_H$=60 GeV) & $\alpha$ ($m_H$=300 GeV)& 
$\alpha$ ($m_H=1$ TeV) \\ \hline
45 & 5.6 & 8.4 & 12.0 \\ \hline 
80 & 2.4 & 3.4 & 4.8 \\ \hline 
250 & 0.56 & 0.76 & 1.0 \\ \hline\hline
\end{tabular}
\end{center}
\caption{Upper bounds on values of $\alpha$ for different triplet and SM 
Higgs boson masses}
\end{table}

\end{document}